# Performance analysis of common browser extensions for cryptojacking detection*

Dmitry Tanana
*Laboratory of Combinatorial Algebra*
*Ural Federal University*
Yekaterinburg, Russia
ddtanana@urfu.ru

***Abstract*—This paper considers five extensions for Chromium-based browsers in order to determine effectiveness of browser-based defenses against cryptojacking, which are available to regular users. We've examined most popular extensions - MinerBlock, AdGuard AdBlocker, Easy Redirect && Prevent Cryptojacking, CoinEater, and Miners Shield, which claim to be designed specifically to identify and prevent illegal cryptocurrency mining. An empirically confirmed dataset of 373 distinct cryptojacking-infected websites, which was assembled during multi-stage procedure, was used to test those extensions. The results showed that all plugins in question had significant performance limits. Easy Redirect and Miners Shield only blocked 6 and 5 websites, respectively, while MinerBlock had the greatest detection rate at 27% (101/373 sites blocked). Most concerningly, despite promises of cryptojacking prevention, AdGuard (which has over 13 million users) and CoinEater were unable to block any of the compromised websites. These results demonstrate serious flaws in cryptojacking detection products targeted for regular users, since even the best-performing specimen failed to detect 73% of attacks. The obvious difference between advertised capabilities and real performance highlights the urgent need for either accessibility improvements for laboratory-grade detection technologies that show more than 90% efficiency in controlled environment or fundamental upgrades to current commonly used extensions.**

## I. INTRODUCTION

The total value of the world's cryptocurrency market in 2024 was over 2.2 trillion dollars [1]. Generating new cryptocoins, which is known as mining, usually demands a huge amount of computational power. For example, Bitcoin, the biggest of all cryptonetworks, generated over 600 million terahashes per second in May 2024 [2] and used more than 100 TWh of electric power in 2023 [3]. Because of their widespread use, cryptocurrencies make a desirable target for many types of cybercriminals. Cryptocurrencies are often employed in the shadow economy for anything from drug trades to ransom demands because of their relative untraceability [4]. However, those characteristics also attract another kind of cybercriminals, who utilize victims' computers without authorization to mine cryptocurrency. These kinds of attacks are referred to as malicious mining or cryptojacking.

There are two primary types of malicious mining attacks [5]:

1. Executable-type cryptojackers, which are often a common mining software that has been altered to include configurable settings by cybercriminals. Via spam, exploit kits, or other malware, it is installed on the victims' machine.
2. Browser-based cryptojackers: they appear in a user's browser when they visit a website that has a mining script running. If a website obtains users approval for mining, it is deemed lawful in the majority of countries.

The Coinhive project, which mined Monero coin, was the first to make browser-based mining widely available. The creators' initial goal was to provide website owners an alternative to using advertisements as a source of revenue, but it soon gained a lot of traction among cybercriminals. Browser-based cryptojacking attacks have significantly dropped once Coinhive was shut down in 2019. but cybercriminals quickly shifted to other mining services [6].

Attacks using cryptojacking are very common in the banking industry these days. A more recent data from SonicWall claims that since the start of 2022, the number of cryptojacking assaults in financial institutions has grown by 269%. In the first half of 2023, there was a 30% rise in the total number of cryptojacking occurrences. Currently, there are five times as many assaults on the financial sector than there were in the second half of 2022 [7].

Cryptojacking has repercussions that go beyond simple resource theft. Because further malware may be introduced by the first infection, it may act as a springboard for more complex assaults. Also, due to decentralized nature of most cryptocurrencies, it is difficult to identify the recipients of these mining operations, which gives cybercriminals even more confidence in their anonymity. Strong detection and prevention systems are essential as cryptocurrencies continue to become more widely used in order to protect users' devices and data.

In this study we conduct a survey and comparative analysis of Chromium extensions that claim to be able to safeguard the user while visiting a website that has been compromised by a cryptojacker. Chromium-based browsers were chosen because of their market share, which is around 80% for 2025. Similar studies can be performed for Firefox and Safari extensions, but given their relatively low user base, we've decided to focus on Chromium-based extensions. The increasing frequency of cryptojacking attacks and the relative insufficiency of existing end-user defenses that are available to regular users are the driving forces behind this study. Client-side technologies like browser extensions are crucial for end users who could inadvertently access infected websites, even when server-side protections are available. This research intends to evaluate these technologies in order to identify weaknesses in current solutions and direct users toward more efficient solutions.

This work was supported by the Ministry of Science and Higher Education of the Russian Federation, project no. FEUZ-2023-0022.

*This paper was presented as part of the 2025 XI International Symposium on Problems of Redundancy in Information and Control Systems (Redundancy) and published in its proceedings

This paper's structure is set up as follows: Relevant papers which consider cryptojacking detection and prevention are reviewed in Section II. The methodology used, including test stand design and infected website set is described in Section III. The result table and concluding thoughts are presented in Section IV, which also discusses future prospects. Finally, a list of references is provided.

## II. Previous Works

The paper "Sok: Cryptojacking malware" by Tekiner et al. [8] focuses on cryptojacking as on one of the most common attack vectors, affecting everything from financial institutions to videoconferencing platforms. The authors note the shortcomings of current detection techniques and how attackers might get around them. In addition to offering insights and research ideas for the academic community interested in this area, the paper suggests a methodical study of the malware based on datasets, attack instances, and academic research in order to solve this issue.

"End-to-End Analysis of In-Browser Cryptojacking" by Saad et al. [9] emphasizes the explosive growth of cryptojacking. Both static and dynamic analysis are used to detect cryptojacking in the authors' thorough investigation. Along with reviewing current countermeasures and their shortcomings, the paper makes recommendations for long-term countermeasures based on the results of the studies conducted.

We also like to highlight the fundamental strategy used by T.P. Khiruparaj et al. in "Unmasking File-Based Cryptojacking" [10], where the authors have developed a mathematical model to assess the likelihood of cryptojacking detection. The authors provide a mathematical methodology for detecting cryptojacking that uses mean CPU load and the interquartile range technique to identify load spikes. They claim a 98% detection rate by combining this with network packet analysis. Nevertheless, the research leaves out important performance measures that are necessary for a thorough assessment of the model's effectiveness, particularly the probabilities of Type I (false positive) and Type II (false negative) mistakes.

In our previous work, "Advanced Behavior-Based Technique for Cryptojacking Malware Detection" [11] we've proposed an advanced behavior-based technique for cryptojacking malware detection. That approach focused on monitoring a specific set of metrics: CPU usage, network activity, and calls to cryptographic libraries in order to identify malicious mining activities. By enhancing previous heuristic-based methods, we've achieved a detection rate of 93% on selected sample set, emphasizing the importance of dynamic analysis over static signatures. This method is effective against both file-based and browser-based cryptojackers, addressing limitations in earlier works by incorporating multi-core CPU monitoring and empirical thresholds for network and library calls. Program developed during that study is used to validate infected website list for this paper.

Furthermore, A. Zareh Chahoki et al. in "Cryptojackingtrap: An evasion resilient nature-inspired algorithm to detect cryptojacking malware" [12] introduce CryptojackingTrap, an evasion-resilient nature-inspired algorithm for detecting cryptojacking malware. Inspired by the Venus Flytrap mechanism, this solution uses dynamic binary instrumentation to correlate memory access traces with cryptocurrency network data, focusing on hash function executions. It supports multiple cryptocurrencies through extensible plugins and achieves zero false negatives and false positives in evaluations, with a mathematically calculated false positive rate of $10^{-20}$. This work advances detection by targeting unavoidable mining behaviors, resistant to obfuscation and encryption techniques.

Together, these papers highlight how cryptojacking risks are always changing and how creative detection techniques are required. Dynamic and behavior-based techniques, as shown in [11] and [12], provide stronger resistance against complex evasion techniques, employed by cryptojackers, while static analysis yield fundamental insights. All of the aforementioned investigations, however, depend on specially designed detection techniques created by scientists. The prototypes are experimental and at best are available in obscure Git repositories and at worst not available at all. This study will assess Chromium-based browser extensions that are available for regular users in the Chrome web store and claim the capacity to identify and prevent cryptojacking.

## III. Methodology

This work focuses on the performance analysis of clientside protection tools, specifically – browser extensions. They were tested on a list of websites which are running cryptojacking scripts, allowing for an evaluation of the detection and/or prevention capabilities in a real-life scenario.

To evaluate the tools, we've adopted an approach of selecting extensions for Google Chrome, since it is the most popular browser in 2025 [13]. The selection process followed two stages: Initially, we conducted a search with the keyword "cryptojacking" in the Chrome store, in order to identify available extensions. At this stage, we obtained a list with only 10 results. Then, we've used a simple classification technique: whether the extension's description specifies cryptojacking detection and/or prevention. Based on these criteria, we've selected five available extensions, which were:

• Easy Redirect && Prevent Cryptojacking (Version 3.7.0): blocks harmful URLs and redirects the user away from infected websites through the use of regular expressions, thus protecting against cryptomining;

• AdGuard AdBlocker (Version 4.3.12): blocks YouTube and Facebook ads; its description also mentions protection against adware, spyware, malware, phishing, and cryptojacking attacks;

• CoinEater (Version 0.1.3): blocks cryptocurrency miners; its block list is claimed to be based on regular internet checks, within a scientific research project by the Institute for IT Security Research at St. Pölten University of Applied Sciences;

• minerBlock (Version 1.2.18): a browser extension that aims to block browser-based cryptocurrency miners across the web. The extension uses two different approaches to block miners. The first is based on blocking requests/scripts loaded from a blacklist, which is the traditional approach adopted by most mining blockers and other adblockers. The other approach supposedly detects potential mining behavior within loaded scripts and eliminates them immediately;

• Miners Shield (Version 0.1.6): monitors the user's CPU consumption data to prevent possible cryptojacking attacks, blocking the sites responsible for the unusual increase in CPU usage.

To test our set of extensions we've used stock Windows 11 11H2 installation on VMWare virtual machine, with Google Chrome (v. 135), running on 32 gb RAM and Intel Core i5-11300h hardware. Each extension was tested independently to prevent cross-interference and powershell script was designed to automate the process.

### A. *Sample collection*

To conduct the experiment with selected Chrome extensions, we've used two popular databases containing websites running cryptojacking scripts. The first set consisted of 3496 URL addresses [14], and the second had 490 URLs [15], which were then combined into a single set with 3986 URLs total.

After that we've used a simple script to test infected server's response. If the server responded with status 200, indicating that the site was active and functioning, we'll count it as active. Out of the 3986 collected URLs, 553 were active.

Finally, we used the standalone cryptojacking detection program, developed as part of "Advanced Behavior-Based Technique for Cryptojacking Malware Detection" [11] to check if those websites were not only active, but actually running cryptojacking scripts. With this method we were able to identify 373 unique infected websites out of 553 active websites.

The process from database acquisition to final validation is shown in Figure 1, which depicts our URL filtering procedure. Because inactive or non-infected websites may distort findings, this multistep URL filtering makes sure that all sample websites in the final list are relevant. Our proven cryptojacking detection technology, which relies on identifying cryptojackers using dynamic analysis is used to confirm the cryptojacking activity.

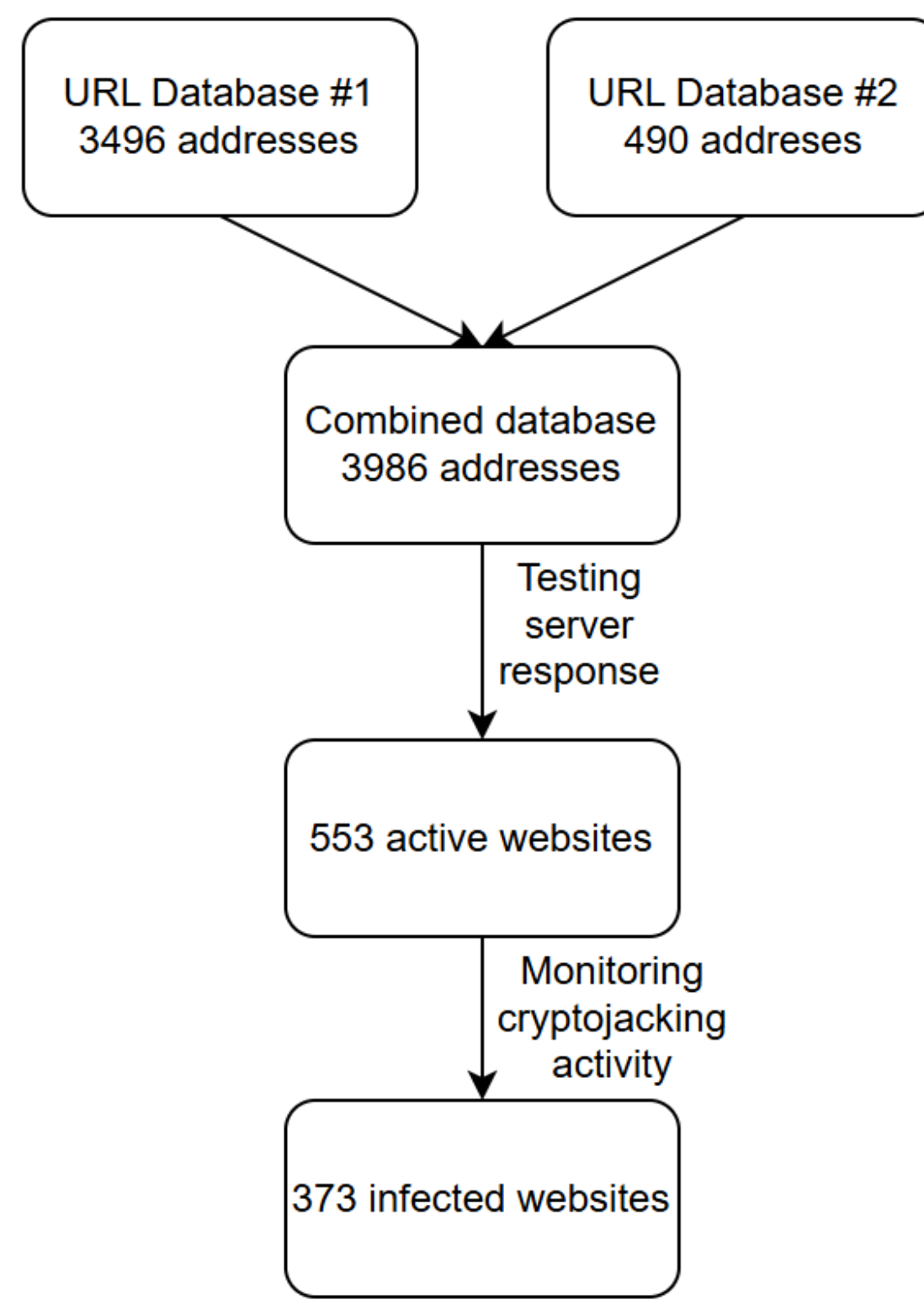

Fig. 1. Infected website filtration process.

The produced list of 373 compromised websites should illustrate a wide range of live threats, with different kinds of cryptojacking scripts running unique evasion strategies. Given that cryptojacking often displays polymorphism to avoid blacklists and static analysis, this variety is essential for evaluating the performance of blocking extensions.

## IV. RESULTS AND CONCLUSION

To verify whether the cryptojacking process on specific website is prevented by each extension, we've created a powershell script which inserted the URLs from our list of 373 infected websites into browser, waited for a minute and then took a screenshot. After that we've manually checked all the screenshots to see if the website loaded and/or alert message was displayed. At the end of the process, we've recorded the websites where extension was able to identify and/or prevent cryptojacking (exact reaction on detected cryptojacking activity depends on extension in question) and the corresponding URLs.

The results are presented in Table I. The Miner Block extension obtained the best performance at 27% blocking rate, managing to block 101 of the 373 infected websites, followed by the Easy Redirect extension and the Miners Shield extension, both with blocking rate around 1%. The AdGuard AdBlocker and Coin Eater extensions were unable to block any of the sites from our list.

TABLE I. NUMBER OF WEBSITES BLOCKED BY EACH EXTENSION

| Extension | Number of users | Blocked websites |
|---|---|---|
| Miner Block | 200 000 | 101 |
| AdGuard AdBlocker | 13 000 000 | 0 |
| Easy Redirect && Prevent Cryptojacking | 2000 | 6 |
| Coin Eater | 893 | 0 |
| Miners Shield | 104 | 5 |

The combined strategy used by Miner Block, blacklist-based blocking in conjunction with behavioral analysis and mining script detection, may be the reason for its better performance. The hybrid approach was demonstrated in other publications, such as [11] and [12], which supports behavior analysis as a means of detecting cryptojackers employing stealth techniques.

On the other hand, even while AdGuard's description makes assertions about preventing cryptojacking, with its emphasis on adblocking, it may not adequately address behavior unique to cryptocurrency mining. Coin Eater's reliance on obtaining blacklist from webserver could explain its ineffectiveness if the blacklist is outdated compared to our URL database.

The regular users require more adaptable tools that include dynamic analysis and/or machine learning, as shown by the poor overall detection rates of anti-cryptojacking browser extensions. It should be noted that these findings were produced in controlled environment and future results may be impacted by real-world variability, such as extension and/or cryptojacker upgrades.

Overall, this paper presented a comparative performance study of the most widely used anti-cryptojacking extensions for the most popular web browser in the world. The findings are concerning: even though the Miner Block extension has the greatest blocking rate, it is less than 30%. Furthermore, out of all the extensions that were evaluated, AdGuard AdBlocker had the most users, counting more than 13 million and had the poorest results, failing to detect and block any infected websites.

It is crucial to remember that the results shown might be affected by the databases selected; they could also change as a result of updates for already-existing extensions or new extensions added to the Chrome store. Both the dynamic nature of cryptojacking threats and the unique detection algorithms of each extension may affect detection and prevention rates.

The results of this study show that the widely available technologies for user defense – simple browser extensions do not adequately handle the majority of cryptojacking threats, highlighting serious flaws in the present security ecosystem. The need for improved detection methods is highlighted by our results, perhaps using behavior-based strategies like those discussed in papers from our Previous Works section, which show consistent effectiveness of more than 90%.

Future research should consider the question of bringing laboratory-grade cryptojacking detection techniques to regular users, for example by creating and marketing a unique browser extension or collaborating with creators of well-known anti-ad extensions.

This research has implications for policy-making and user education, promoting a better understanding of the current technology limits and the dangers of cryptojacking.

## REFERENCES

[1] R. de Best, "Overall Cryptocurrency Market Capitalization Per Week from July 2010 to July 2024", Available at: https://www.statista.com/statistics/730876/cryptocurrency-maket-value/.

[2] "Blockchain.com | Charts – Total Hash Rate", Available at: https://www.blockchain.com/en/explorer/charts/hash-rate

[3] A. de Vries, "Bitcoin Energy Consumption Index", Available at: https://digiconomist.net/bitcoin-energy-consumption/.

[4] S. Kethineni, Y. Cao, "The Rise in Popularity of Cryptocurrency and Associated Criminal Activity", International Criminal Justice Review, vol. 30, February 2019, pp. 325–344.

[5] S. Varlioglu, N. Elsayed, Z. ElSayed and M. Ozer, "The Dangerous Combo: Fileless Malware and Cryptojacking," Proceedings of SoutheastCon 2022, Mobile, AL, USA, 2022, pp. 125-132

[6] Varlioglu S., Gonen B., Ozer M., Bastug M. F. "Is Cryptojacking Dead After Coinhive Shutdown?", Proceedings of the 3rd International Conference on Information and Computer Technologies, 2020, pp. 385–389.

[7] "Sonicwall cyber threat report", Available at: https://www.sonicwall.com/cyber-threat-report/.

[8] E. Tekiner, A. Acar, A. S. Uluagac, E. Kirda and A. A. Selcuk, "SoK: Cryptojacking Malware," 2021 IEEE European Symposium on Security and Privacy (EuroS&P), Vienna, Austria, 2021, pp. 120-139.

[9] M. Saad, A. Khormali, and A. Mohaisen, "End-to-End Analysis of In-Browser Cryptojacking", 2018, Available at: http://arxiv.org/abs/1809.02152.

[10] T.P. Khiruparaj, V. Abishek Madhu, P. R. K. Sathia Bhama, "Unmasking File-Based Cryptojacking", Advances in Intelligent Systems and Computing, Vol. 1167, December 2019, pp. 137-146.

[11] D. Tanana, G. Tanana, "Advanced Behavior-Based Technique for Cryptojacking Malware Detection," 2020 14th International Conference on Signal Processing and Communication Systems (ICSPCS), Adelaide, SA, Australia, 2020, pp. 1-4.

[12] A. Zareh Chahoki, H. R. Shahriari and M. Roveri, "CryptojackingTrap: An Evasion Resilient Nature-Inspired Algorithm to Detect Cryptojacking Malware," in IEEE Transactions on Information Forensics and Security, vol. 19, pp. 7465-7477, 2024.

[13] "Browser Statistics 2025", Available at: https://www.w3schools.com/browsers/.

[14] "CoinBlockersList", Available at: https://zerodot1.gitlab.io/CoinBlockerListsWeb/.

[15] "Nefarious Laboratories Blocklist", Available at: https://github.com/mmg1/blocklists/tree/master.